\documentclass[a4paper]{jpconf}
\usepackage{graphicx,iopams}
\begin{document}
\title{Phase diagram of asymmetric Fermi gas across Feshbach resonance}

\author{C.-H. Pao$^1$ and S.-K. Yip$^2$}

\address{$^1$Department of Physics, National Chung Cheng University, Chiayi 621,
Taiwan}
\address{$^2$Institute of Physics, Academia
Sinica, Nankang, Taipei 115, Taiwan}
\ead{pao@phy.ccu.edu.tw}

\begin{abstract}
We study the phase diagram of the dilute two-component Fermi gas at zero temperature as a function of the polarization and coupling strength. We map out the detailed phase separations between superfluid and normal states near the Feshbach resonance. We show that there are three different coexistence of superfluid and normal phases corresponding to phase separated states between: (I) the partially polarized superfluid and the fully polarized normal phases, (II) the unpolarized superfluid and the fully polarized normal phases and (III) the unpolarized superfluid and the partially polarized normal phases from strong-coupling BEC side to weak-coupling BCS side. For pairing between two species, we found this phase separation regime gets wider and moves toward the BEC side for the majority species are heavier but shifts to BCS side and becomes narrow if they are lighter.
\end{abstract}

\section{\label{intro}Introduction:}

Recently experiments in the ultracold Fermi gases\cite{fermions}
have raised strong interest in studying the crossover problem from
Bose-Einstein condensation (BEC) to the condensation of
weak-coupling Cooper pairs. The experiments
\cite{ketterle06,hulet06} on $^6$Li atoms with imbalance spin
populations further provide an opportunity to study the superfluid
properties with mismatched Fermi surface. The phase diagram of this
polarized Fermi system has been studied near the Feshbach resonance
for both of the pairing with equal masses \cite{pao06,sheehy06} and
different species \cite{Iskin06,lin06,wu06,pao07,Parish}. The homogeneous
superfluid is stable only at the large positive coupling strengths.
There are two possible scenarios to replace the unstable phases,
namely phase separation \cite{Sarma} and FFLO states \cite{FFLO}.
 Determining phase separation is relatively
straight-forward.  On the other hand, calculations for the FFLO
states are highly non-trivial due to the complex spatial dependences
of the order parameter \cite{yip07}.  However, the instability of
the normal state towards the FFLO state is relatively easy to
determine if one assumes a second order transition, for then one
only needs to examine the Cooperon at finite q. In this paper, we
shall then construct the phase diagram for phase separation using a
Maxwell construction \cite{paoyip06}, but then terminating the
procedure whenever the corresponding normal state has a finite-q Cooperon instability.  The latter regions are presumably in the
FFLO state.

For the same species, we found that there are three different
coexistence of superfluid and normal phases corresponding to the
phase separation between (I) the partially polarized superfluid
(${\rm SF_P}$) and the fully polarized normal (${\rm N_{FP}}$)
phases, (II) the unpolarized superfluid (${\rm SF_0}$) and the fully
polarized normal phases, and (III) the fully polarized superfluid
and the partially polarized normal (${\rm N_{PP}}$) states.

For the pairing between different species, this phase separation
region becomes wider compared to the pairing of equal masses for the
case of the majority species are heavier but narrower if the
majority species are lighter. For the case with $\gamma \equiv
m_{majority} /m_{minority} = 6.67$ (between $^{40}$K and $^6$Li),
the instability of the finite-q Cooperon
occurs earlier than the instability of the phase separation for almost all couplings
so that there are effectively only two kinds of phase
separation (${\rm SF_P + \rm N_{FP}}$ and ${\rm SF_0 + N_{FP}}$).
For the other case with $\gamma = 0.15$, the phase separation region
shifts to BCS side and all of three kinds of phase separated states
still exist.

\section{\label{form}Formalism:}
We start from the two-component fermion system across a wide Feshbach resonance
which may be described by an effective one-channel Hamiltonian as follows:
\begin{equation}
H\ =\ \label{eqh} \sum_{{\bf k}, \sigma} \xi_\sigma ({\bf k})
c^\dagger_{{\bf k},\sigma} c_{{\bf k},\sigma}\ +\ g \sum_{\bf
k,k^\prime, q} c^\dagger_{{\bf k+q},h} c^\dagger_{{\bf
k^\prime -q},l} c_{{\bf k^\prime},l} c_{{\bf k},h}\ ,
\end{equation}
where $\xi_{\sigma}({\bf k})\, =\, \hbar^2 k^2/2m_\sigma -
\mu_\sigma$, $g$ is the bare coupling strength, and the index
$\sigma$ runs over the two species ($h$ and $l$), where $-h \equiv
l$.  $m_h$ ($m_l$) to represent the mass of the  heavier (lighter)
component.  Within the BCS mean field approximation, the excitation
spectrum in a homogeneous system for each species is
\begin{equation}
E_{h,l}({\bf k})\ =\ \mp \left [ { \hbar^2 k^2 \over 4 m_r} { \gamma
-1 \over \gamma + 1}\, +\, h \right ]\ +\ \sqrt{ \left ({\hbar^2 k^2
\over 4 m_r}\, -\, \mu \right )^2 \, +\, \Delta^2 }\ ,
\label{eqdisp1}
\end{equation}
with the reduced mass $m_r$, the chemical potential
difference $h\, \equiv (\mu_h - \mu_l)/2$ and the average chemical potential
$\mu\, \equiv\, (\mu_h + \mu_l)/2$.

It is convenient to re-scale the energy in unit of the chemical
potential difference $h$, then the equation for pairing field
$\Delta$ reads
\begin{equation}
{1 \over 2 \pi \tilde{a}}\Delta \ =\ -\Delta \int{ d^3 \tilde{k} \over ( 2 \pi)^3}
\left [ { 1 - f( E_h / |h|) -f(E_l /|h|) \over (E_h + E_l)/|h| }\ -\ { 2 \over {\tilde{k}}^2} \right ]\ , \label{gapeq}
\end{equation}
where $\tilde{k} = k/\sqrt{m_r |h|}$, $\tilde{a} = a \sqrt{m_r |h|} $, and $\hbar =1$. The total density ($N$) and the polarization ($P$) of the system are
\begin{eqnarray}
N & =& N_h + N_l\ =\ (m_r |h|)^{3/2} \int { d^3 \tilde{k} \over ( 2 \pi)^3} \left [
1 \, +\, 2 \left ( {\tilde{k}^2 \over 4} - \tilde{\mu} \right ) {f( E_h / |h|) + f(E_l /|h|)-1 \over (E_h + E_l)/|h| } \right ]\ ,\\
 P & = & {N_h - N_l \over N} \ =\ {(m_r |h|)^{3/2}\over N} \int { d^3 \tilde{k} \over ( 2 \pi)^3} \biggl [ f( E_h / |h|) -f(E_l /|h|) \biggr ]\ ,
\end{eqnarray}
with $\tilde{\mu} = \mu / |h|$. For a given $\tilde{\mu}$, the scaled
scattering length $\tilde{a}$ can be evaluated as a function of the
pairing field $\Delta/|h|$ through equation (\ref{gapeq}). When there
are multiple solutions at the same $\mu$ and $h$,  the physical
solution is then determined by the condition of minimum free energy
which can be found via the same procedure as the usual Maxwell
construction \cite{paoyip06}.

\section{\label{result}Results and Discussions}

\begin{figure}[h]
\includegraphics[width=16pc]{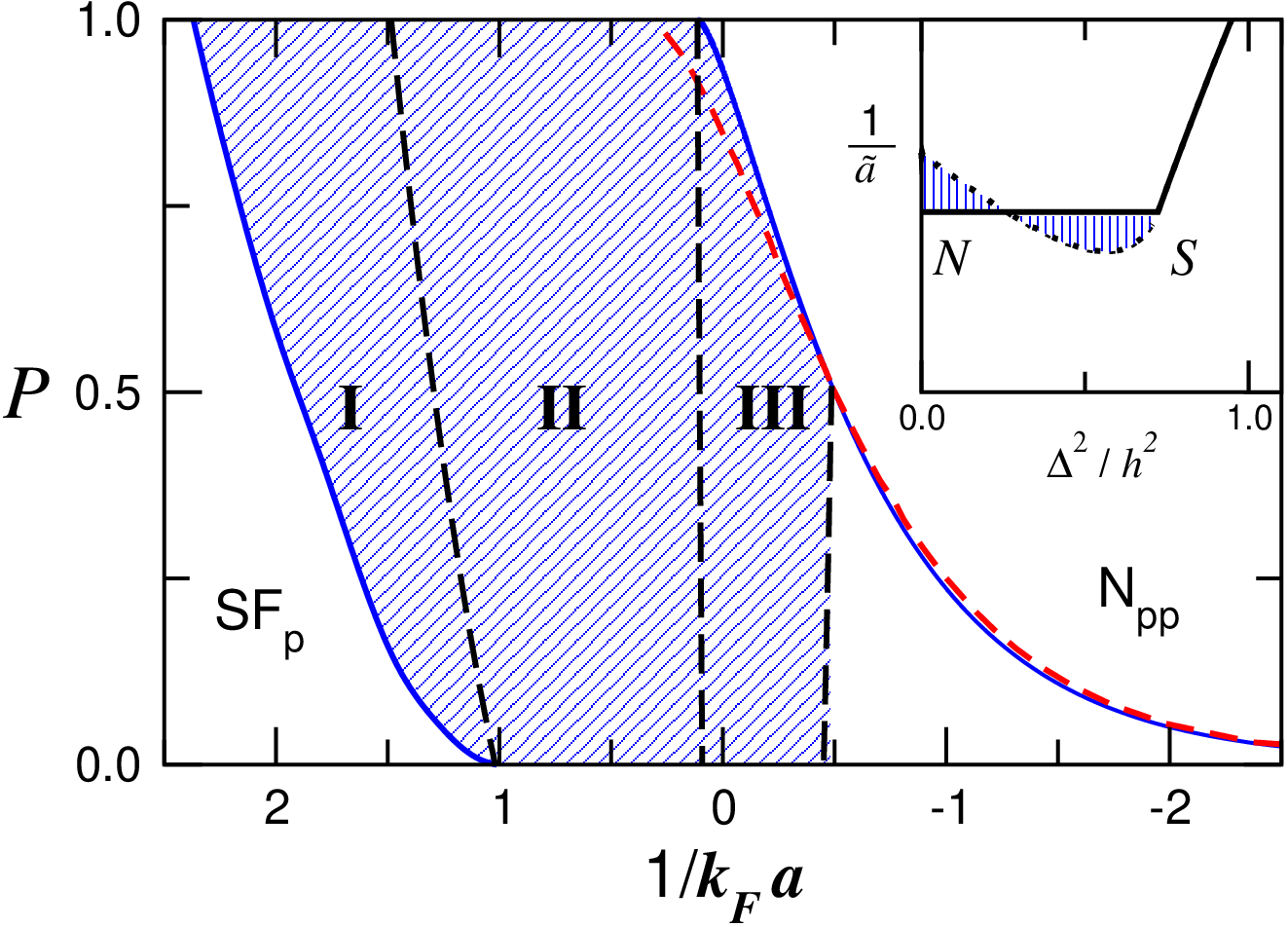}\hspace{2pc}%
\begin{minipage}[b]{19pc}\caption{\label{fig1} Phase diagram for the imbalanced fermion system with $\gamma = 1.0$. The homogeneous states (labelled as ${\rm SF_P}$ and ${\rm N_{PP}}$) are stable only at large positive or negative coupling strengths. The shaded area represents three different phase separated states (see text). The short dashed lines represents the instability of the finite-$q$ Cooperon. The inset shows a typical solution of equation (\ref{gapeq}) at fixed $\tilde{\mu}$ in the phase separation region. The shaded areas above and below the horizontal line are equal.}
\end{minipage}
\end{figure}
In Fig. \ref{fig1}, we plot the phase diagram of the polarization as
a function of the coupling strength [$(k_F a)^{-1}$] for the pairing
with equal masses. The Fermi momentum ($k_F = (3 \pi^2 N)^{1/3}$) is
the usual definition for the unpolarized non-interacting Fermi gas.
In a polarized system, the homogeneous state is stable only at large
positive and negative coupling strengths  corresponding to the
partially polarized superfluid (BEC side) and the partially
polarized normal state (BCS side). The shaded
area near the resonance ($|a| \rightarrow \infty$) represents the
phase separation between superfluid fluid and normal gas
\cite{pao06,sheehy06}. Furthermore, this shaded area is divided into
three regions: the phase separation between (I) ${\rm SF_p\, and \,
N_{FP}}$, (II) ${\rm SF_0\, and \, N_{FP}}$, and (III) ${\rm SF_0\,
and \, N_{PP}}$. (This division is slightly different from the Monte
Carlo results \cite{Pilati}). The phase boundaries among these three
regions are determined by the Maxwell construction at fixed
$\tilde{\mu}$ and $\tilde{a}$. A typical example is shown in the
inset in figure \ref{fig1}. The two end points (labelled as $N$ and
$S$) of the horizontal line are corresponding
to the normal ($\Delta = 0$) and superfluid ($\Delta \ne 0$) states.
Then the phase boundaries (long dashed lines shown in figure
\ref{fig1}) are evaluated by
\begin{eqnarray}
 P_x & = & {x N_d^S + (1-x) N_d^N \over x N^S + (1-x) N^N}\ ,\\
 \left ({ 1 \over k_F a } \right )_x& = & { \sqrt{m_r h} / \tilde{a} \over [ 3 \pi^2 (x N^S + (1-x) N^N) ]^{1/3}}\ ,
 \end{eqnarray}
 where $N^{S,N}= N^{S,N}_h + N^{S,N}_l$, $N^{S,N}_d=N^{S,N}_h - N^{S,N}_l $, and $0 \le x \le 1$.
  As increasing $\tilde{\mu}$, we will reach the BCS regime, the instabilities of the phase separation and finite $q$ Cooperon intersect near $P \sim 0.5$ and $(k_F a)^{-1} \sim -0.49$. The FFLO phase probably exists on the right of this point \cite{yip07}.

\begin{figure}[h]
\begin{minipage}{18pc}
\includegraphics[width=16pc]{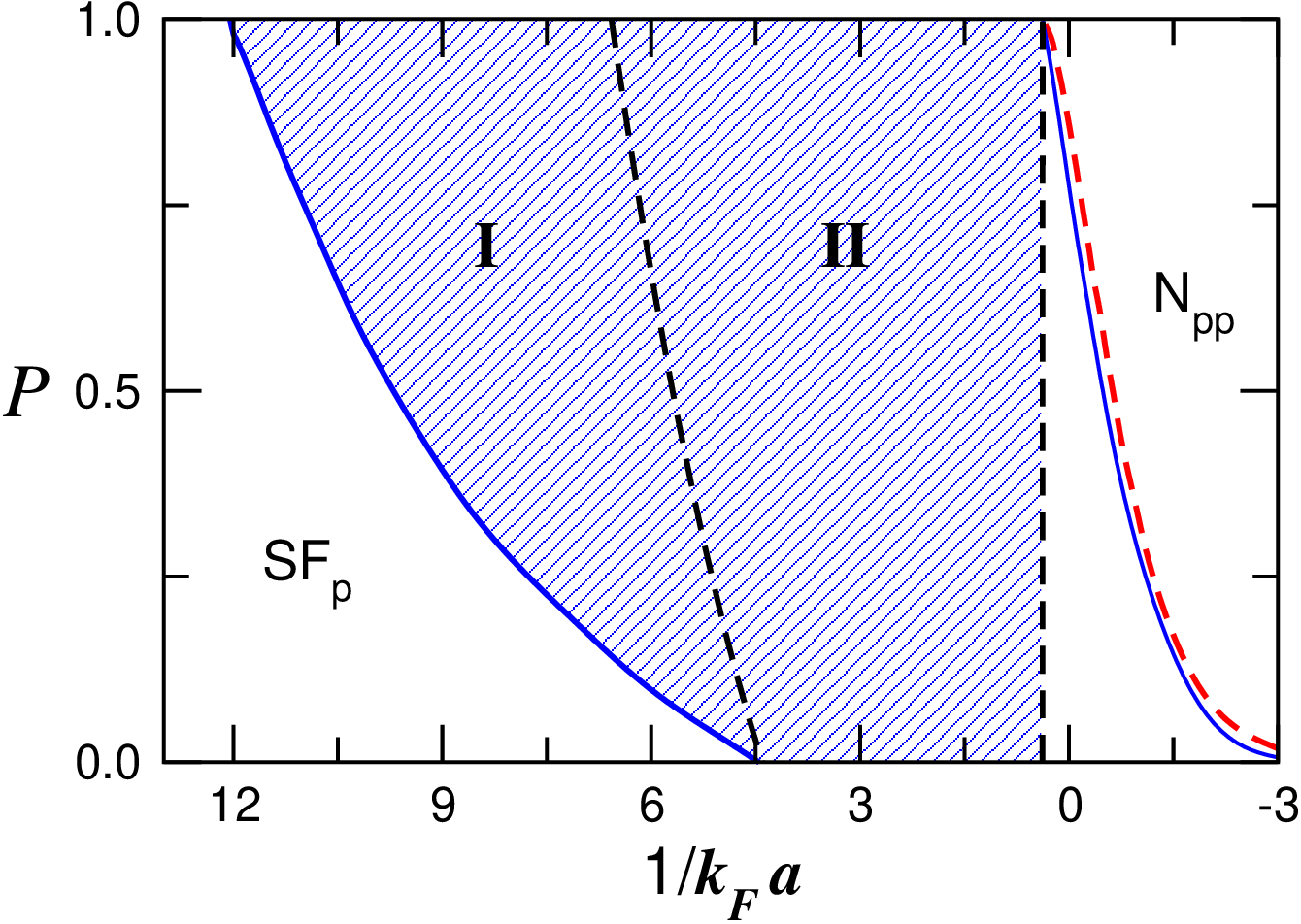}
\caption{\label{fig2}Phase diagram for the imbalanced fermion system with
 $\gamma = 6.67$. Notations are the same as in figure \ref{fig1} but only two kinds of phase separated states exist here (see text).}
\end{minipage}\hspace{2pc}%
\begin{minipage}{18pc}
\includegraphics[width=16pc]{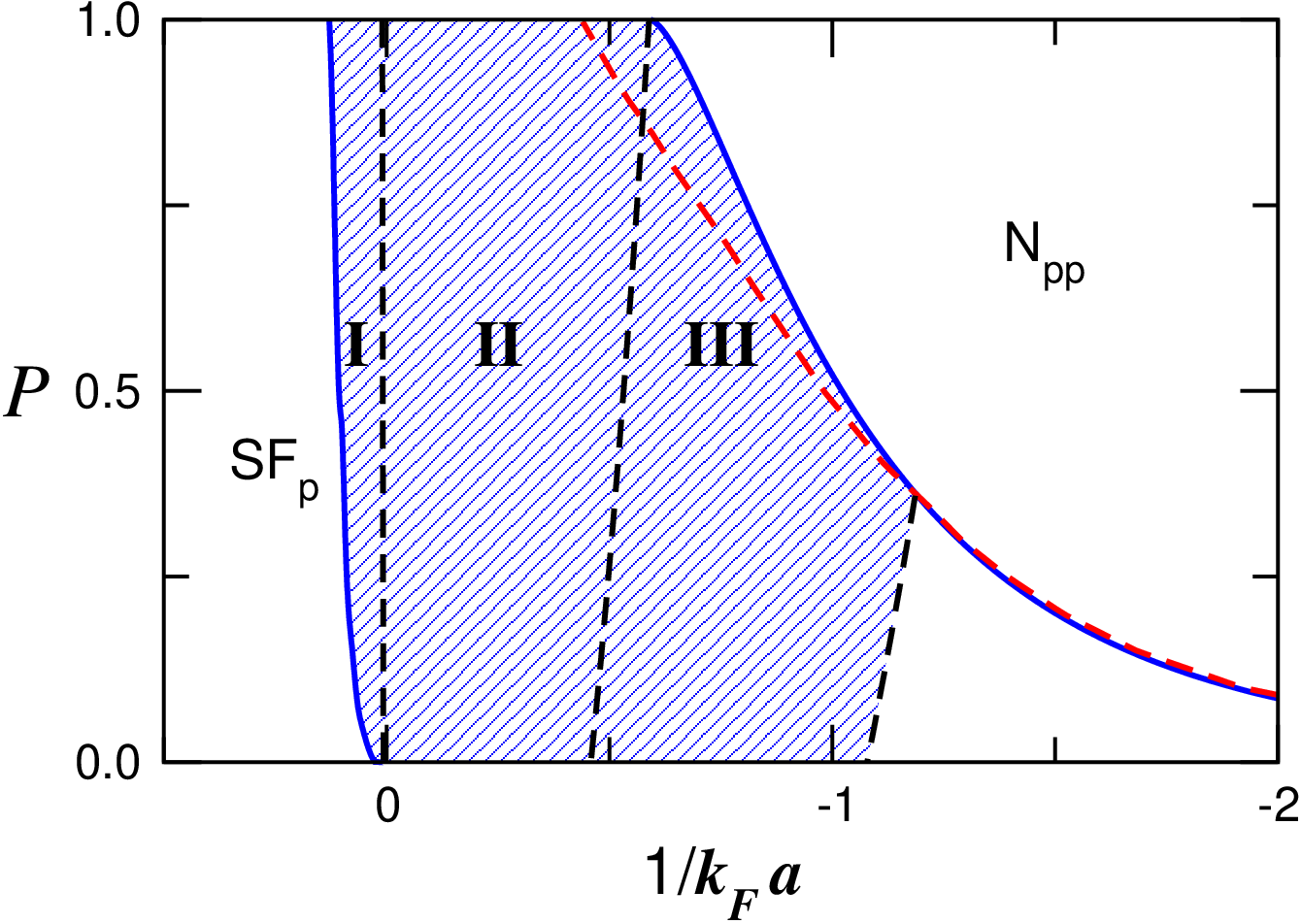}
\caption{\label{fig3}Phase diagram for the imbalanced fermion system with
 $\gamma = 0.15$. Notations are the same as in figure \ref{fig1}.\vspace*{0.15in}}
\end{minipage}
\end{figure}

For the pairing between different species, the story is quite different. For the case of the majority species are heavier, the phase separation region moves toward the BEC side and becomes much wider than the case of equal masses pairing. In figure \ref{fig2}, we plot the phase diagram for $\gamma = 6.67$. The instability of the finite-$q$ Cooperon occurs earlier compared to the instability of the phase separation for almost entire polarization such that the phase separation between ${\rm SF_0\, and\, N_{PP}}$ effectively does not exist in this case.

When the majority species are lighter, the phase separation region gets narrow and shifts toward the BCS side. In figure \ref{fig3}, we plot the phase diagram for $\gamma = 0.15$. As increasing the coupling strengths from BCS limit, the system prefers entering FFLO state first for $ P \lesssim 0.36$ but the phase separated states between the ${\rm SF_0\, and\, N_{PP}}$ for $ P \gtrsim 0.36$. All of the three phase separated states exist in this case.

\section{\label{conclusion}Conclusion}

We have studied the phase diagram for the pairing between two species in a dilute Fermi gas at zero temperature. Near the Feshbach resonance, we map out the detailed phase separations between superfluid and normal state. There are three different phase separated states for the paring of equal masses. For the majority species are lighter, the phase separation region becomes narrow and shifts to BCS side compared to the case of equal masses. On the other hand, this region gets wider and only the phase separated states of ${\rm SF_p + N_{FP}}$ and ${\rm SF_0 + N_{FP}}$ exist for $\gamma = 6.67$.

\ack This research was supported by the National Science Council of
Taiwan under grant numbers NSC96-2112-M-194-007-MY3 (CHP) and
NSC96-2112-M-001-054-MY3 (SKY), with additional support from
National Center for Theoretical Sciences, Hsinchu, Taiwan.


\section{References}

\end{document}